\documentclass[proceedings, preprint]{rmaa}

\usepackage{paralist}

\usepackage{psfrag,color}

\SetYear{2007}
\SetConfTitle{The Nuclear Region, Host Galaxy and Environment of Active Galaxies}

\title{Observational Overview of the Feeding of Active Galactic Nuclei} 

\author{Thaisa Storchi-Bergmann\altaffilmark{1}}

\altaffiltext{1}{Instituto de F\'{i}isica, UFRGS, Campus do Vale,
CP15051, Porto Alegre, RS, Brazil (thaisa@ufrgs.br).}

\shortauthor{Storchi-Bergmann}
\shorttitle{Feeding of AGN}

\listofauthors{T. Storchi-Bergmann}
\indexauthor{Storchi-Bergmann, T.}
\abstract{I present an overview of the observational signatures of
feeding of Active Galactic Nuclei, discussing briefly the role of interactions among galaxies on extragalactic scales, and of non-axisymmetric gravitational potentials -- such as bars -- on galactic scales. Then I discuss at larger length 
the feeding signatures on hundred of parsec scales, for which new results include:
(1) recent star formation surrounding the active nucleus on tens of parsec scales; 
(2) excess of  gas and dust in active galaxies relative to non-active ones, 
in the form of nuclear spirals and disks; (3) new kinematic signatures of
gas inflow along nuclear spiral arms, which may be the long sought mechanism
to bring gas from kiloparsec scales down to the nucleus to feed the supermassive black hole.}

\addkeyword{Active Galactic Nuclei}
\addkeyword{Black Holes}
\addkeyword{Galaxies}
\addkeyword{Stellar Population}
\addkeyword{Gas kinematics}

\begin{document}
% Typeset article header
\maketitle

\section{Introduction}
\label{sec:intro}

Over the last decade we have learned, mainly from observations
with the Hubble Space Telescope, that most -- if not all -- galaxies which have a stellar bulge, harbor a Supermassive Black Hole (hereafter SMBH) in their 
nuclei, with a mass proportional to that of the bulge  
(e.g. Magorrian et al. 1998; Tremaine et al. 2002; Marconi \& Hunt 2003). 
The difference between an active and a non-active galaxy in this scenario is the 
fact that, in the former, the SMBH is actively accreting
mass, while in the non-active galaxies, the SMBH is ``starving''.

The main unsolved questions regarding the
nuclear activity in galaxies are the origin of the accreting
mass to feed the SMBH and the nature of the process which triggers 
the nuclear activity, including the mechanisms to bring
gas from galactic scales to the nucleus of the galaxy (Martini 2004).

Proposed mechanisms to promote gas inflow to the center of galaxies 
consist, on extragalactic scales, mainly of interactions with other galaxies (Hernquist 1989; Barnes \& Hernquist 1992).
Then, inside the galaxies, non-axisymmetric kiloparsec scale structures --
such as bars -- can bring gas from the galaxy disk to 
$\sim$ 1\,kpc from the nucleus (e.g.
Shlosman et al. 1990; Shlosman et al. 2000; Combes 2004). On hundred of parsec scales,
nuclear bars (e.g. Englmaier \& Shlosman 2004)  and nuclear spirals  
have been proposed as 
means to bring gas from one kiloparsec inwards (Pogge \& Martini 2002; 
Maciejewski 2004a, 2004b). 

%Finaly, on sub-pc scales, accretion disks
%are the proposed means to feed the SMBH (Sakura \& Sunyaev 1973;
%Collin ***; Narayan 2000***).

In this overview, I will discuss observational signatures of the processes
outlined above, which include relevant contributions by Deborah and collaborators.

\section{Feeding on extragalactic and galactic scales}

The feeding on the largest scales -- extragalactic and galactic -- 
is extensively revised by I. M\'arquez, in this proceedings, thus I will
just briefly discuss some personal perspective based mostly on my previous
works and on those of my collaborators in this subject.

As mentioned in the introduction, theoretical studies
propose interactions among galaxies as efficient means to promote gas
inflow to the central region of galaxies. The observational studies by Deborah and
collaborators partially support these theoretical studies as they find
and excess of close ($\le$100\,kpc) companions in Seyfert 2
galaxies in comparison with control samples of non-active galaxies
(Dultzin-Hacyan et al. 1999; Koulouridis et al. 2006a). They also find that there is no excess of
companions among Seyfert 1 galaxies, suggesting a difference
between Seyfert 1 and Seyfert 2 galaxies which seems to contradict the Unified Model (Antonucci 1993).

On the other hand, observations using different criteria to charaterize the
environment, have  failed to find an excess of  companions in samples of 
active galaxies when compared with control samples of non-active ones
(e.g. Schmitt 2001, Li \& Kauffmann 2006). 

In Storchi-Bergmann et al. (2001), where we looked for signatures of interactions among a sample of 35 Seyfert galaxies, we found an excess of companions only for the sub-sample of 
active galaxies which showed an excess of young to intermediate 
age (10$^8$\,yrs) stars in the stellar population. The active galaxies 
dominated by older stellar populations did not show an excess of companions.
Storchi-Bergmann et al. (2001) then proposed an evolutionary scenario,
in which the interactions were the first step in sending gas inwards.
This gas once accumulated in the central regions, would then trigger
both star-formation and the nuclear activity. There are a few examples
of these composite nuclei (e.g. Gonzalez Delgado et al. 1998), because the starburst fades quickly, 
in just a few Myrs, while the nuclear activity lasts at least ten times longer. 
Alternatively, the nuclear activity can be triggered after the 
starburst; the gas to feed the SMBH could come from the mass loss from
the evolving stars. The result is that,
during most of the active phase, clear signatures of interactions
are mostly gone; they are only present shortly after the triggering of the starburst, which corresponds to a short fraction of the active phase.
A somewhat similar evolutionary scenario has been proposed by 
Koulouridis et al. (2006b), in which they find that
bright IRAS galaxies show even more signs of close interaction
than they had found for Seyfert 2 galaxies, as well as lots of
recent star formation. They propose that these galaxies evolve to Seyfert 2 galaxies when the star formation rate decreases and the signatures of interaction get weaker (e.g. companions move away) and the Seyfert 2
galaxies then evolve to Seyfert 1.

On galactic scales, Shlosman, Combes and collaborators have been claiming that bars are efficient means to 
promote gas flows to the nuclear region (e.g. Combes 2004). Regarding the observations, there is some controversy: while some groups do not find excess of 
bars in active galaxies samples when compared with control samples
(e.g. Mulchaey \& Regan 1997) others claim that there is such an excess
(Laine et al. 2002). Although inward flows have indeed been found along
bars (e.g. Mundell et al. 1999; NUGA survey: e.g. Garcia-Burillo et al. 2005; Boone et al. 2007), theory and even observations
show that large scale bars many times end at the Inner Lindblad 
Ressonance, located at hundred or parsecs from the
nuclei, where it is observed, in many galaxies, a gas rich ring
of recent star-formation. There is probably also a delay between the
flow along the bar and the final step, not yet well understood,
through which the gas is transferred from the circumnuclear ring
to the nucleus and trigger the nuclear activity. Such a delay,
and a possible destruction of the bar in the meantime or during the
active phase could explain the non-clear association between the presence of
bars and nuclear activity.

\section{Feeding on hundred of pc scales: Stellar Population}

If large ammounts of gas are sent to the central region of galaxies, most 
probably this gas will form stars in its way in. Perry \& Dyson (1985) 
were the first to propose a ``Starburst-AGN connection" (hereafter 
we will use the usual notation AGN to mean Active Galactic Nuclei) in which the
mass-loss from stars evolving in a circumnuclear starburst would provide
the fuel for the SMBH and even for the formation of the broad line clouds.
Then Terlevich and collaborators (e.g. Terlevich \& Melnick 1985; Terlevich et al. 1995) explored this idea further, proposing
that the blue light observed around the nuclei of Seyfert 2 galaxies
was due to young starbursts (Cid Fernandes \& Terlevich 1995). Norman \& 
Scoville (1988) proposed that nuclear starbursts could evolve to AGN, while 
Collin \& Zahn (1999) proposed that star formation could occur in the 
outskirsts of the accretion disk. A similar proposition was put forth by
Wada \& Norman (2002), showing that starburst could hide within the 
obscuring tori surrounding AGN (Antonucci 1993).

Regarding observations, Storchi-Bergmann et al. (2000) and Gonz\'alez Delgado et al. (1998, 2001) have found features in AGN spectra indicative
of the presence of young stars, while Nelson \& Whittle (1996) and Oliva et al. (1996) have argued that AGN hosts have lower mass-to-light ratio than non-active 
galaxies, what can also be explained by the presence of a systematically
younger stellar population.

Our group, using the technique of spectral synthesis (e.g. Cid Fernandes et al. 2001, 2005) have also found an excess of young to intermediate age (10$^6$--10$^8$\,yr) stars in Seyfert galaxies samples when compared with control samples of non-active galaxies. While in Seyfert samples we could find very young starbursts (Schmitt et al. 1999; Storchi-Bergmann et al. 2000; 
Gonz\'alez Delgado et al. 2001), 
in lower activity galaxies and in radio-galaxies we seldom found
starbursts younger than 10$^8$\,yr. We have studied the stellar population in  a radio galaxy sample as compared with a control sample of elliptical galaxies and found that the main difference between the two samples was
an excess of 10$^9$\,yrs old stellar population observed in the radio
galaxies (Raimann et al. 2004). This was also found for a sample of
low-activity galaxies (Gonz\'alez Delgado et al. 2004; Cid Fernandes et al. 2004).
These results could be interpreted within the evolutionary scenario
discussed in the previous section as due to a larger delay between the
onset of a burst of star-formation (possibly due to an interaction sending
gas inwards) and the nuclear activity in radio-galaxies and 
low-activity galaxies, although part of the latter galaxies could
also be evolved stages of the nuclear activity in which there is less
fuel available for the SMBH.

Serote-Ross et al. (1998) and Boisson et al. (2000, 2004) also find 
excess signatures of recent star formation in the circumnuclear regions
of Seyfert 2 galaxies when compared with non-Seyfert samples (see also Gu, Dultzyin-Hacyan \& De Diego 2001). Kauffmann et al. (2003), by studying the stellar population
of thousand of active galaxies in the SLOAN Digital Sky Survey also find
a relation between starburst and nuclear acitivity: the most luminous Seyfert galaxies in their sample present the largest contribution from young stars in the circumnuclear region.

Intermediate age starbursts have also been found in radio-galaxies (e.g. Aretxaga et al. (2001)  and QSO's (Canalizzo \& Stockton, 2001;
Canalizo et al. 2006).

\subsection{Stellar population on tens of parsecs scales}

\begin{figure}[!t]
  \includegraphics[width=1\columnwidth]{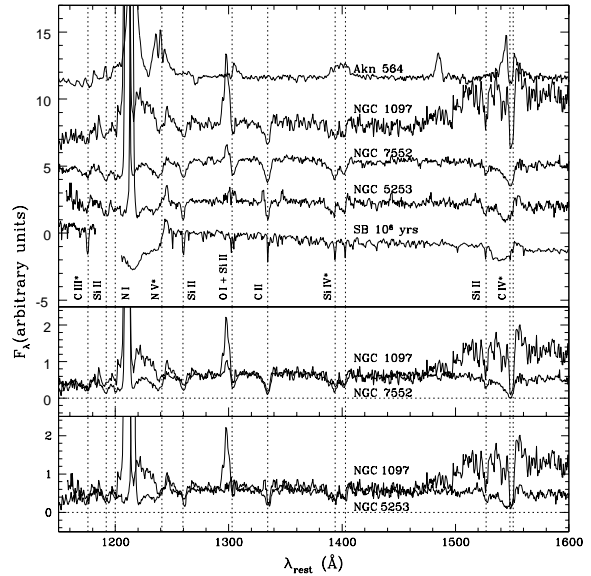}
  \caption{HST UV spectrum of the LINER nucleus of NGC\,1097 as compared with (from to 
to bottom) the spectrum of the Seyfert 1 galaxy Akn\,564, two starburst nuclei and the model of a 10$^6$ yr old starburst, showing the absorption features typical of young O and B stars (from Storchi-Bergmann et al. 2005).}
  \label{UVspec}
\end{figure}

Using HST-STIS spectra, Storchi-Bergmann et al. (2005) have found
absorption lines from O and B stars in the nuclear spectrum of
the LINER/Seyfert 1 galaxy NGC\,1097, within an aperture corresponding 
to a projected distance of only 9\,pc at the galaxy. Fig.\ref{UVspec} shows
the nuclear UV spectrum of NGC\,1097 as compared with those of starburst
galaxies, showing absorption lines originated in the atmospheres of O and B stars, evidencing that the spectral features are characteristic of
a starburst.

\begin{figure}[!t]
  \includegraphics[width=1\columnwidth]{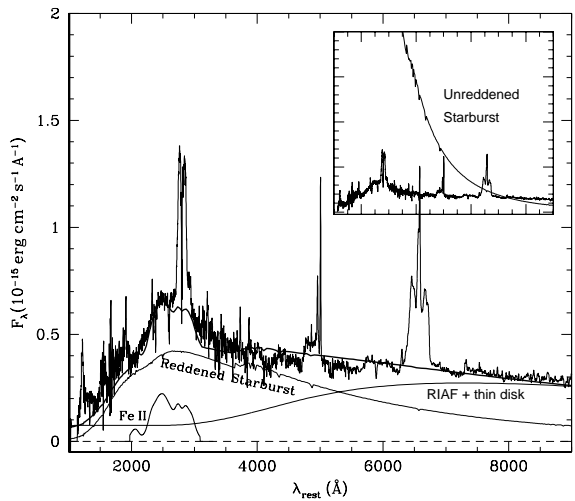}
  \caption{Nuclear spectrum of NGC\,1097, well fitted by a combination of the spectrum 
of a starburst plus a RIAF (radiativelly inneficient accretion flow) model and FeII model emission. The inset shows the unreddened spectrum of the starburst compared to the NGC\,1097 nuclear spectrum.}
\label{sbcont}
\end{figure}

Fig. \ref{sbcont} shouws that,
besides the absorption features, the continuum is also reproduced
by the spectral distribution of a young starburst, confirming the interpretation
based on the UV spectrum. In fact, the continuum shows an odd shape,
with flux increasing towards the blue, as expected for a young starburst, but 
turning down for wavelengths smaller than $\sim$2500\AA. We found out that this  shape could be reproduced by the continuum of a young starburst, once it was
reddened by a reddening law similar to that of the Small Magellanic Cloud for A$_V$=3\,mag. 

The combined UV and optical spectrum of NGC\,1097 illustrated in Figs. \ref{UVspec} and \ref{sbcont} was extracted from the Atlas
of AGN spectra of Spinelli et al. (2006), where many more similar AGN spectra  can be found, which can be recovered via internet. Description of these spectra and how to recover them can be found in the paper above.

Signatures of compact starbursts (sizes $\le$50\,pc) have been recently found around the nuclei of nearby AGNs by Davies et al. (2007) using two-dimensional spectroscopy with the instrument SINFONI at the VLT. The estimated ages of the starbursts are
50-100\,Myr. Davies et al. also map the stellar velocity dispersions $\sigma_\star$, and 
find $\sigma_\star$-drops delineating a compact stellar disk. These disks of low $\sigma_\star$
have been previously reported by Emsellem et al. (2001) \& M\'arquez et al. (2003), among others, and have been interpreted by Emsellem et al. (2001) as due to a young stellar population born from a dynamically cold gas component
(see also Emsellem 2006).

\section{Feeding on hundred of pc scales: morphology}

\begin{figure}[!t]
  \includegraphics[angle=-90,width=1\columnwidth]{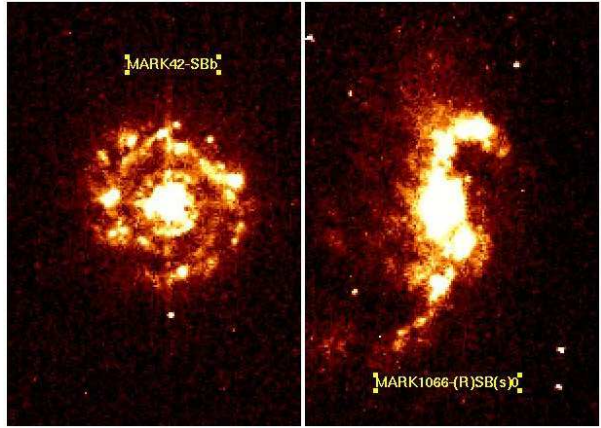}
  \caption{Near-UV HST F303W images of the Seyfert galaxies Mrk\,42 (left), 
which shows a a circumnuclear ring of star formation and Mrk\,1066 (right), which
shows elongated and diffuse emission.}
\label{ACS1}
\end{figure}

\noindent{\it Morphology in the Near-UV.} In our investigation of the stellar population, many times the spectral synthesis 
showed a degeneracy between the youngest stellar population component and a power-law
used to represent the direct or scattered non-stellar continuum (Cid Fernandes et al. 2001). In order to further investigate the nature of the UV light in active galaxies,
we (P.I. H. R. Schmitt) have proposed an HST-ACS snapshot survey of Seyfert galaxies through which we have obtained 75 near-UV images, which have
been collected in an Atlas (Mu\~noz Mar\'{i}n et al. 2007). We expected the morphology to
help revealing the nature of the near-UV light. Fig.\ref{ACS1} shows two typical cases:
in the left, the near-UV image of the galaxy Mrk\,42 shows a clear ring of star-formation surrounding the nucleus, while in the right, the galaxy Mrk\,1066 shows elongated and more diffuse UV emission whose nature is not clear from the UV image alone. We are now in the
process of comparing the UV image with those in other bands to further investigate
the nature of the UV light in the cases similar to that of Mrk\,1066, where there are no obvious structures which are clearly due to recent star formation.

\noindent{\it Morphology in the optical.} Many studies using HST
optical images have pointed out that active galaxies shows a lot of dusty structures within the inner few hundred of parsecs. Van Dokkum \& Franx (1995) have
shown that radio-loud early-type galaxies have more dust than radio-quiet. Pogge \& Martini (2002) and Martini et al. (2003) have shown that most Seyfert galaxies
present dusty filaments and spirals in the nuclear region. Xilouris \& Papadakis (2002) have concluded that, among early Hubble types, active galaxies present more dusty structures than non-active; similarly, Lauer et al. (2006) have argued that the
presence of dust in early-type galaxies is correlated with nuclear activity. Regarding the dusty structures in early-type galaxies, Ferrarese et al. (2006) find signatures of recent star formation in the most regular and compact dust structures.

\begin{figure}[t]
\centering

\begin{minipage}{1\linewidth}
\includegraphics[scale=0.35]{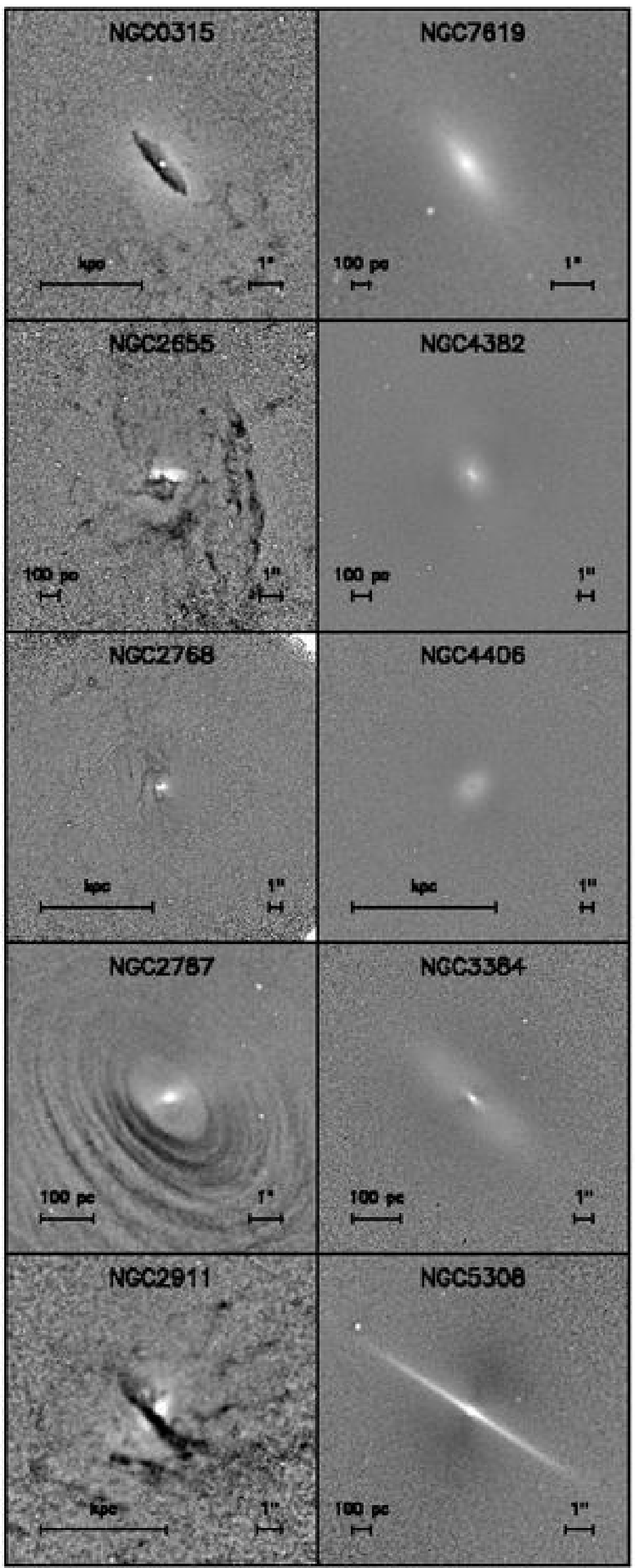}
\includegraphics[scale=0.35]{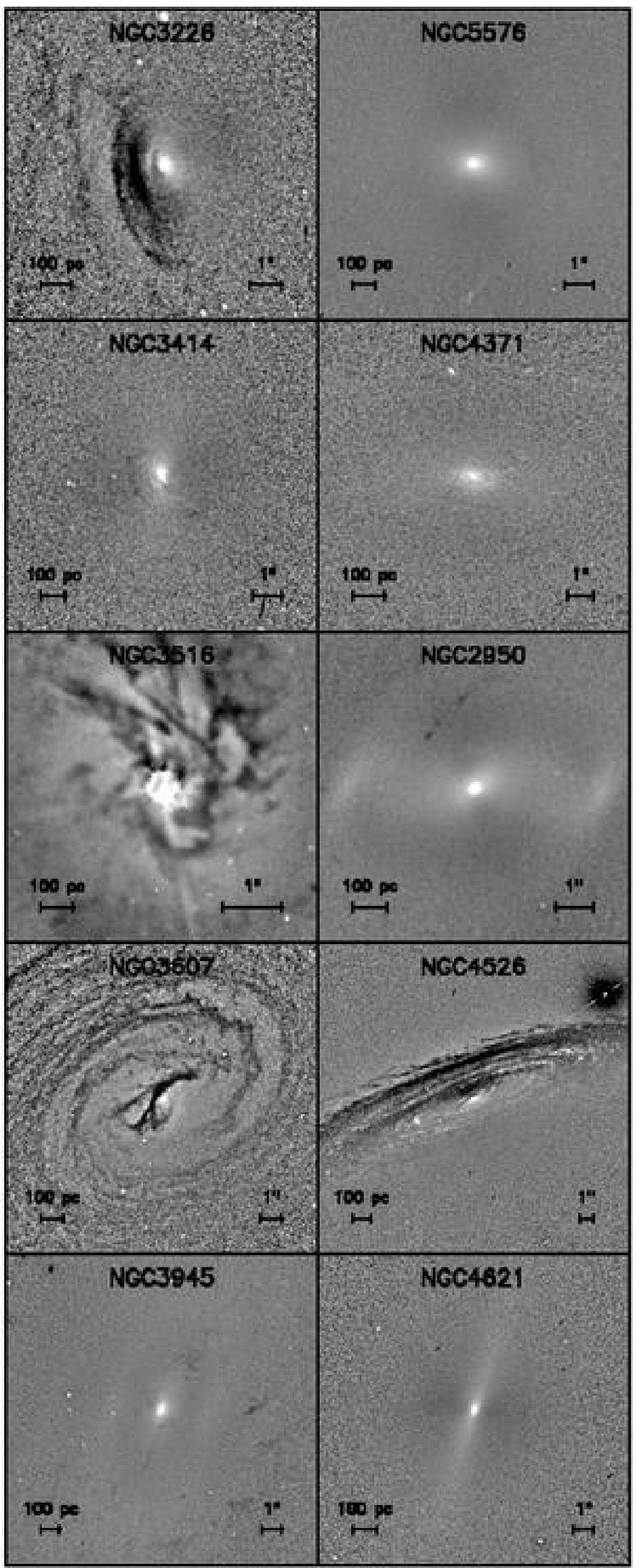}
\end{minipage}

%\begin{minipage}{1\linewidth}
%\includegraphics[scale=0.3]{struct1.eps}
%\includegraphics[scale=0.3]{struct2.eps}
%\end{minipage}

\caption{Structure maps of the nuclear region of 10 pairs of active (left column) and non-active (right column) early-type galaxies, showing that all active galaxies have dusty filamentary/spiral structures while only a few of the non-active galaxies present such structures (from Sim\~oes Lopes et al. 2007).}
\label{struct}
\end{figure}

In order to quantify the association between nuclear activity and the presence of gas and dust in the nuclear region, we (Sim\~oes Lopes et al 2007) have assembled a sample of active galaxies and a control sample of non-active galaxies matched according to the properties of the host galaxies. We have thus constructed matched pairs of galaxies such that each member of the pair could be considered the same galaxy in two phases, one active and the other non-active. A large sample was extracted from the Palomar survey of nearby galaxies (Ho et al. 1995) and from that sample, we were able to find HST optical images for 68  matched pairs of galaxies. In order to compare the structures present in each pair we used the structure map technique proposed by Pogge \& Martini (2002) applied to the HST images. The technique improves the contrast of the images
enhancing structures as fine as the scale of the point spread function, which corresponds at the galaxies to scales of a few pc to a few tens of pc. The results of this study for the 34 early-type pairs of the sample are illustrated in Fig.\ref{struct}: dust structures are found in all active galaxies, but in only 27\% of the non-active galaxies, which represents a strong correlation between the presence of dust structures and activity in galaxies. The dust traces the presence of gas also, and our interpretation of the above correlation is that the dust is tracing the actual material in its way in to feed the SMBH.

An unexpected result of the above study was the finding that at least 50\% of non-active galaxies present nuclear stellar disks, which are not seen in the active galaxies, and which seem to be somewhat more compact than the dust structures in the active galaxies. These stellar disks in early-type galaxies have been reported by previous authors. Krajnovic \& Jaffe (2004), for example, argue that these disks are probably the result of the infall of mass to the center of the galaxy driven by the secular  evolution of a bar, galaxy mergers, or both. 

The dust structures seen in the active galaxies present a range of morphologies, from chaotic filaments, through more organized spiral structure to compact disks, which led Lauer et al. (2006) to suggest evolution or a ``settling sequence''. The stellar disks seen in the non-active galaxies add one more final step in the sequence, leading to the following possible evolutionary scenario: the matter arrives to the inner few hundred parsecs of the galaxy in a chaotic morphology, and then it gradually settles towards the center in more organized structures, such as spirals and dusty disks. As the gas falls in, it feeds the SMBH, leading to nuclear activity. The gas most probably forms also some stars, giving origin to the stellar disks, which are only visible when the dust and gas have been all consumed by the SMBH and/or have been blown away by winds of evolving stars. 

As stellar disks are long-lived, most probably they are also present in the active phase, but are hidden by the large amounts of dust associated with the gas flowing in. Acording to this scenario, during each activity cycle, the stellar disk is replenished with gas which gives rise to star formation in the disk. After the star formation ceases and the gas is all consumed by the SMBH and/or has been blown away by the evolving starburst, the stellar disk is unveiled.

The presence of nuclear stellar disks surrounding  AGNs has been confirmed by the work of Davies and collaborators (e.g. Davies et al. 2007; see also contribution by Mueller Sanchez, in this conference), using the instrument SINFONI (a near-infrared Integral Field Unit at VLT): they map the stellar velocity dispersion and find, surrounding a number of nearby AGNs, compact stellar disks ($\sim$50\,pc) with low velocity dispersions, interpreted as stars formed from cold material that has fallen to the nuclear region.

\section{Feeding on hundred of pc scales: gas kinematics}

It is well known that the kinematics of gas surrounding AGN is dominated by outflows (e.g. Creenshaw et al. 2003 and references therein; Creenshaw et al. 2006). Thus although we believe gas is reaching the center to feed the SMBH, inflows are seldom observed, because they become overwhelmed by the dominating outflows. 

Motivated by the results reported in the previous section of a strong correlation between activity and the presence of dusty nuclear spirals, suggesting that these spirals are channels to feed the SMBH, we have begun a project to measure the kinematics of these spirals using the integral field unit of the Gemini Multi-Object Spectrograph (GMOS IFU). The first results of this project were published in Fathi et al. (2006) for the galaxy NGC\,1097, in which we mapped the H$\alpha$ and [NII]$\lambda$6584 velocity fields within the inner kiloparsec of the galaxy. Although the kinematics is dominated by rotation in a plane, the subtraction of an exponential rotating disk model fitted to the data revealed residuals which traced the spiral arms seen both in an HST image and in near-IR images (Prieto et al. 2005). Notice that the nucleus of NGC\,1097 is a LINER, with low activity, and this is probably the reason why we can see the inflows: because the outflows are weak or absent. 

\begin{figure}[!t]
  \includegraphics[width=1\columnwidth]{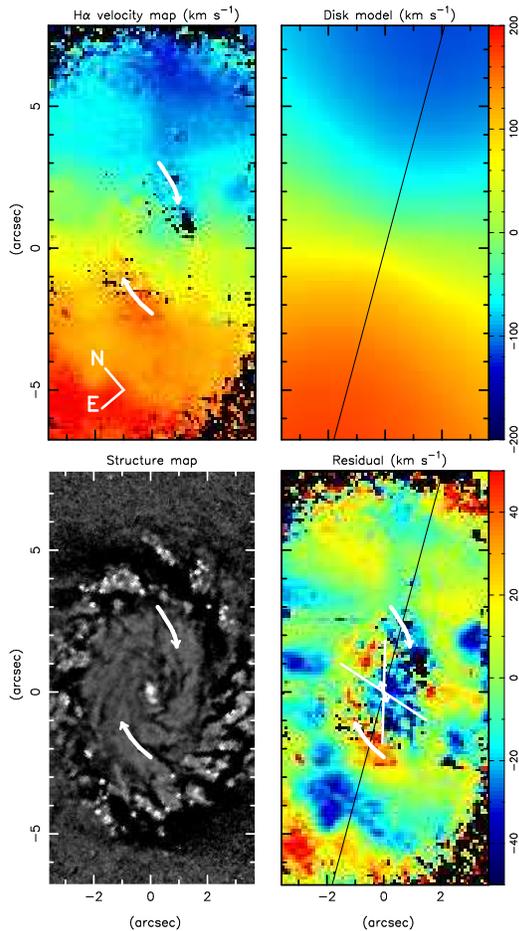}
  \caption{Top left: the H$\alpha$ velocity field. Top right:
model velocity field. Bottom left: structure map.
Bottom Right: difference between observed and model velocity fields.
Curved lines with arrows identify two velocity structures 
which deviate from the model: the ``W arm''\,(top) and 
the ``E arm''\,(bottom). The small bar with arrows at the
nucleus in the bottom right panel represents a compact radio source (Saikia et al. 2002),
and the two white lines delineate a bi-cone within which we observe
an enhancement in the vellocity dispersion due to the interaction of the
radio jet with the circumnuclear gas.
The straight black line shows the location of the line of nodes.}
\label{kin}
\end{figure}

We have recently found similar results to those described above for NGC\,1097 around another LINER nucleus, NGC\,6951 (Storchi-Bergmann et al. 2007), which are shown in Fig. \ref{kin}. In the top left panel we show the velocity obtained from the peak wavelength of the H$\alpha$ emission line, where arrows show deviations from circular rotation along two partial spiral arms. An exponential disk model, illustrated in the top right panel of this figure was again subtracted from the observed velocity field, leading to the residuals shown in the bottom right panel. The residuals include non-circular motions along the star-forming ring shown in the structure map in the bottom left panel of the figure, and, within the ring, residuals due to two motions: streaming motions along the partial spiral arms identified in the top left panel combined with outflows due to the effect of a nuclear radio jet in the interstellar gas. We have discovered the effect of the radio jet in the velocity dispersion map, which showed enhancements in the gas velocity dispersion in two blobs to NW and SE, approximately within 1 arcsecond from the nucleus. Then we found out that there is indeed an elongated radio source observed by Saikia et al. (2002), extended in the direction where we observed the enhancements in the velocity dispersion, in agreement with the interpretation that the interaction of the radio jet  with the circumnuclear ISM is producing the observed outflows. Thus, in this LINER, we could observe both the inflows and outflows. We are now observing more low activity nuclei with nuclear spirals to investigate if these inflows are confirmed in most cases and to constrain the mass inflow rate.

These inflows, now observed in two cases, suggest that we are witnessing, for the first time, the mass going to the nuclear region to feed the SMBH.
Such streaming motions had been observed previously only on large scale spiral arms (e.g. Emsellem et al. 2006). Using the typical streaming velocities (40-50\,km\,s$^{-1}$) and the observed geometry for the flow from the images, as well as using estimates for the filling factor, it is possible to estimate the inflow rate in ionized gas, which gives values of approximately 10$^{-3}$solar masses per year in both cases (NGC\,1097 and NGC\,6951), and which coincides with the accretion rate necessary to power both low activity nuclei. Nevertheless, the inflow in ionized gas is probably only the ``tip of the iceberg''; larger amounts of colder and molecular gas should be flowing in as well. This is supported by the observations of $\sim$10$^7$ solar masses of molecular gas by Garcia Burillo et al. (2005) and Krips et al. (2007) within tens of parsecs of the nucleus of NGC\,6951.

\section{Conclusions}

I have reviewed the observational signatures of feeding processes to active galactic nuclei, from extragalactic scales to hundred of parsec scales.

On extragalactic scales, the favored mechanism is interactions among galaxies, although the observational signatures are not very clear. This seems to be due to a delay between the interaction and most of the active phase of up to several hundred Myr. This interpretation is supported by: (1) the observation that signatures of interaction are seen only for AGNs living in galaxies with signatures of recent circumnuclear star-formation; (2) the presence of an excess of intermediate age stellar population in most active galaxies. Thus the age of the last generation of stars probably dates the interaction, which triggers both the star-formation and nuclear activity. The starburst then fades in a few Myrs, while the AGN outlives the starburst; alternatively, there may be also a delay in the onset of nuclear activity relative to the starburst activity.

 On galactic scales, bars seem to be the preferred mechanism, and inflows of neutral and
molecular gas along bars have been observed, although the association of the presence of bars with nuclear activity is not clear or at least controversial, at the moment. One possibility is that there is also a delay between the inflow along the bar and the final inflow on hundred of parsec scales which will feed the SMBH, and that, in the meantime, some bars may disappear.

On hundred of parsec scales, we have found a strong association between the presence of nuclear gaseous spirals and filaments and nuclear activity, which can only be understood if this gas is the actual fuel on its way in to feed the SMBH at the nucleus. This is confirmed by new results reported here of observations of inflows along nuclear spirals, observed for the first time, thanks to the two-dimensional coverage, high image quality and well suited spectral resolution provided by the the Integral Field Units of the Gemini Multi-Object Spectrographs. These measurements have also allowed an estimate of the inflow rate of mass towards the nucleus.

We may have found the long-sought physical mechanism to bring gas inwards to feed the SMBH after the gas reaches the inner kiloparsec of the galaxy: nulcear gaseous spirals, which seem to be shocks in nuclear gaseous disks, which can account for  the loss of angular momentum and allow the inflow of matter to the nucleus. 

In summary, considering: (1) the dominance of intermediate age stellar population in AGN host galaxies; (2) the lack of clear association of AGN activity with with interactions and bars; (3) the association of nuclear spirals with the AGN activity; (4) the presence of nuclear stellar disks both in active and non-active phase, a simple evolutionary scenario for feeding can be:

\begin{itemize}
\item
Interactions among galaxies promote capture of gas by the galaxies;
\item
Non-axisymmetric potentials, such as bars bring gas to $\sim$ 1\,kpc from the nucleus;
\item
Nuclear spirals develop inside the inner kiloparsec and bring gas to the inner tens of parsecs;
\item
The accumulation of gas triggers a starburst within tens of parsecs from the nucleus;
\item
This gas can either immediatly trigger the nuclear activity (leading to the observed composite Seyfert plus Starburst nuclei) or trigger first the starburst activity and mass loss from the evolving stars provide the fuel for the AGN;
\item
The starburst fades to an intermediate age stellar population and the AGN then shines for tens to hundreds of Myrs.
\end{itemize}

I would like to acknowledge the organizers of the Conference for the invitation to present this work, Deborah, for her wonderful character and for being a role model for women in science, as well as for her valuable contributions to Astronomy. I would like to thank also my collaborators, in particular my post-doc (O. Dors Jr.) and students: R. A. Riffel, R. Nemmen, P. Spinelli, R. Sim\~oes Lopes, H. Brandt and F. K. B. Barbosa.

\end{document}